\documentstyle[12pt,psfig,cite]{article}
\begin{document}
\renewcommand{\thefootnote}{\fnsymbol{footnote}}
\begin{titlepage}
\begin{flushright}
KEK-TH-561
\end{flushright}
\vspace*{10mm}
\Large
\begin{center}
Common Structures in Simplicial Quantum Gravity
\end{center}
\vspace{5mm}
\normalsize
\begin{center}
H.S.Egawa  \footnote[2]{E-mail address: egawah@theory.kek.jp} 
, 
N.Tsuda,   \footnote[3]{E-mail address: ntsuda@theory.kek.jp}
and
T.Yukawa   \footnote[5]{E-mail address: yukawa@theory.kek.jp}
\end{center}
\begin{center}
$^{\dag}$ Department of Physics, Tokai University\\
Hiratsuka, Kanagawa 259-12, Japan

$^{\ddag}$ Theory Division, Institute of Particle and Nuclear Studies, \\
KEK, High Energy Accelerator Research Organization\\
Tsukuba, Ibaraki 305 , Japan

$^{\P}$ Coordination Center for Research and Education,\\   
The Graduate University for Advanced Studies,\\
Hayama-cho, Miuragun, Kanagawa 240-01, Japan 
\end{center}
\vspace{5cm}
\parindent 5cm
\begin{abstract}
The statistical properties of dynamically triangulated manifolds 
(DT mfds) in terms of the geodesic distance have been studied  numerically.
The string susceptibility exponents for the boundary surfaces 
in three-dimensional DT mfds were measured numerically.
For spherical boundary surfaces, we obtained a result 
consistent with the case of a two-dimensional spherical DT surface 
described by the matrix model. 
This gives a correct method to reconstruct two-dimensional random 
surfaces from three-dimensional DT mfds. 
Furthermore, a scaling property of the volume distribution of 
minimum neck baby universes was investigated numerically 
in the case of three and four dimensions, and we obtain 
a common scaling structure near to the critical points 
belonging to the strong coupling phase in both dimensions. 
We have evidence for the existence of a common fractal structure 
in three- and four-dimensional simplicial quantum gravity.
\end{abstract}
\end{titlepage}
  
\section{Introduction}
There has been, over the last few years, remarkable progress in the 
quantum theory of two-dimensional gravity.
Two distinct analytic approaches for quantizing two-dimensional gravity 
have been established.
These are recognized as a discretized theory\cite{matrix} and a 
continuous\cite{Liouville} theory.
The discretized approach, implemented by the matrix model technique, 
exhibits behavior found in the continuous approach, given by Liouville 
field theory, in a continuum limit, for example, the string 
susceptibility exponent and Green's function.
Strong evidence thus seems to exist for the equivalence between 
the two theories in two dimensions. 
A numerical method based on the matrix model, such as that of 
dynamical triangulation\cite{dt}, has drawn much attention 
as an alternative approach to studying non-perturbative effects, 
also being capable of handling those cases where analytical 
theories cannot yet produce meaningful results. 
In the dynamical-triangulation method, calculations of 
the partition function are performed by replacing the path integral 
over the metric to a sum over possible triangulations.
Over the past few years a considerable number of numerical studies 
have been made on three- and four-dimensional simplicial quantum 
gravity\cite{3DQG}, and recent results obtained by dynamical 
triangulation in three and four dimensions suggest the existences 
of scaling behaviors near to the critical point
\cite{3D_Scaling,4D_Scaling} in terms of the geodesic distance 
\cite{fractal}.
When the coupling strength ($\kappa$) becomes close to the critical point 
from the strong-coupling side ($\kappa < \kappa^{c}$), 
which corresponds to the crumple phase, the transition is smooth.
On the other hand, when $\kappa$ becomes close to the critical point 
from the weak-coupling side ($\kappa > \kappa^{c}$), which corresponds 
to the branched-polymer phase, the transition is very rapid.
In fact, in ref.\cite{3D_Scaling} it is reported that near to 
the critical point belonging to the strong-coupling phase 
($\kappa \to \kappa_{c}$ from $\kappa < \kappa_{c}$) 
the scaling behavior of the mother boundary surfaces exists 
at appropriate geodesic distances, and also that no mother boundary 
surface exists in the weak-coupling phase 
(i.e., branched polymer phase). 
It seems reasonable to suppose that the model makes sense 
as long as we are close to the critical point from 
the strong-coupling phase.
More noteworthy is that there are several pieces of evidence that the 
phase transition of three- and four-dimensional simplicial gravity is 
first order \cite{3D_1st,4D_1st}.
We actually observe double-peak histogram structures which are signals 
of a first-order phase transition for appropriate large lattice sizes.
Therefore, we carefully chose one of the peaks belonging to the strong 
phase as an ensemble for our simulations.

We must thus look more carefully into these boundary mfds.
In the last few years, several articles have been devoted to 
the study of boundary mfds.
In two dimensions, it is revealed by ref.\cite{IK} that 
the dynamics of the string world sheet (random surfaces) 
can be described by the time (i.e., geodesic distances) 
evolution of boundary loops.
Ref.\cite{SS} is also based on the idea that the functional 
integral for three- and four-dimensional quantum gravity can be 
represented as a superposition of a less-complicated theory of 
random-surface summation over a two-dimensional surface
\footnote{In this case a definition of a time direction is different 
from ours.}.
It is precisely on such grounds that we claim that the higher 
dimensional complicated theory of quantum gravity can be reduced 
to the lower dimensional quantum gravity.

This paper is organized as follows.
In Sec.2 we briefly review the model of the three-dimensional dynamical 
triangulation.
In Sec.3 we report our measurements for the string susceptibility 
exponent ($\gamma_{st}$) of the mother boundary surfaces in three-dimensional 
DT mfds.
In Sec.4 we report on our measurements for the string susceptibility 
exponent ($\gamma_{st}$) of the three- and four-dimensional DT mfds.
We summarize our results and discuss some future problems in the final 
section.
%

\section{Model}
We start with the Euclidean Einstein-Hilbert action 
in $D(=3,4)$ dimensions,  
\begin{equation} 
S_{EH} = \int d^{D} x \sqrt{g} \left(\Lambda - \frac{1}{G}R \right), 
\end{equation}
where $\Lambda$ is the cosmological constant and $G$ is 
Newton's constant of gravity.
We use the lattice action of the $D$-dimensional model 
with the $S^{D}$ topology, corresponding to the above action, 
as follows: 
\begin{eqnarray} 
S(\kappa_{0},\kappa_{D}) & = & - \kappa_{0} N_{0} + \kappa_{D} N_{D}  
          \nonumber\\    
  & = & - \frac{2\pi}{G}N_{0} 
        + \left(
          \Lambda' - \frac{1}{G} 
          (2\pi - 6 \mbox{cos}^{-1}(\frac{1}{D}))
          \right) N_{D}, 
\end{eqnarray}
where $N_i$ denotes the total number of $i$-simplexes, and 
$\Lambda' = c\Lambda$; $c$ is the unit volume of a $D$-simplexes, 
and $\mbox{cos}^{-1}(\frac{1}{D})$ is the angle between two tetrahedra.
The coupling ($\kappa_{0}$)\footnote{In the case of four dimensions, 
we use $\kappa_{2}$ in all runs instead of $\kappa_{0}$ with the 
conventions.} is proportional to the inverse of the bare Newton's 
constant, and the coupling ($\kappa_{D}$) corresponds 
to a lattice cosmological.

For the dynamical triangulation model of $D$-dimensional 
quantum gravity we consider a partition function of the form 
\begin{equation} 
Z(\kappa_{0}, \kappa_{D}) 
= \sum_{T(S^{D})} e^{-S(\kappa_{0}, \kappa_{D})}.
\label{eq:PF} 
\end{equation}
We sum over all simplicial triangulations ($T(S^{D})$) on a 
$D$-dimensional sphere.
In practice, we must add a small correction term ($\Delta S$) 
to the lattice action in order to suppress any volume fluctuations. 
The correction term is denoted by  
\begin{equation} 
\Delta S = \delta (N_{D} - N_{D}^{(\mbox{target})})^2,
\end{equation}
where $N_{D}^{(\mbox{target})}$ is the target volume of 
$D$-simplexes, and we use $\delta = 0.0005$ in all runs.
%

\section{Boundary structure of three-dimensional DT mfds}
%
\begin{figure}
\centerline{\psfig{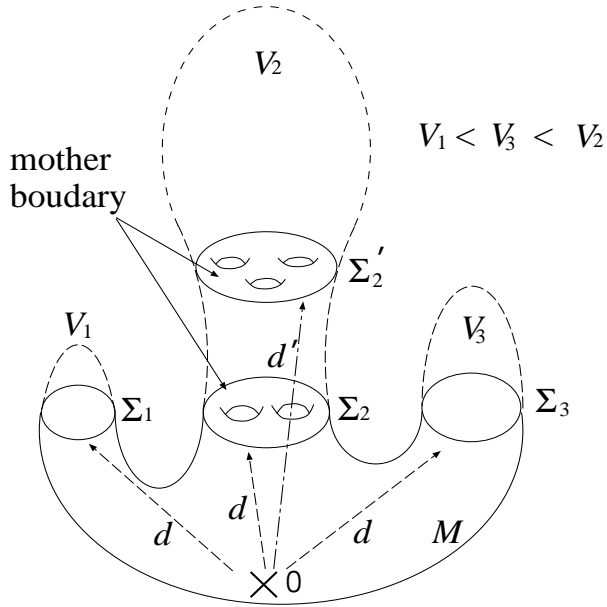}} 
\caption
{
Schematic picture of boundary surfaces ($\Sigma_{1},\Sigma_{2}$ and 
$\Sigma_{3}$) at distance $d$ and a surface ($\Sigma_{2}'$) at distance 
$d'$ in 3D Euclidean space $M$ with $S^{3}$ topology.
The mother boundary surface is defined as a surface ($\Sigma_{2}$) with the 
largest tip volume ($V_{2}$), and the other surfaces are defined as the baby 
boundary.
}
\label{Fig:Boundary_3D_CS}
\end{figure}

%
We now define the intrinsic geometry using the concept of a geodesic 
distance as a minimum length (i.e., minimum step) in the dual lattice 
between two 3-simplexes in a three dimensional DT mfd.
Suppose a three-dimensional ball (3-ball) which is covered within $d$ 
steps from a reference 3-simplex in the three-dimensional mfd with 
$S^{3}$ topology. 
Naively, the 3-ball has a boundary with spherical topology ($S^{2}$).
However, because of the branching of DT mfds, the boundary is not 
always simply-connected, and there usually appear many boundaries which 
consist of closed and orientable two-dimensional surfaces with any 
topology and nontrivial structures, such as links or knots
\footnote{In the strict sense links or knots are constructed by loops. 
In our case the loop is a fat loop.}.
Below, we consider only three-dimensional DT mfds with $S^{3}$ 
topology.
We can show a sketch of typical configuration 
in Fig.\ref{Fig:Boundary_3D_CS}.
$M$ denotes a three-dimensional DT mfds with $S^{3}$ topology and 
$\Sigma_{1}$, $\Sigma_{2}$ and $\Sigma_{3}$ denote the boundaries 
which are closed and orientable triangulated surfaces at a distance 
$d$ from an origin ($\times$).
In a previous paper \cite{RS_3D_ET} we showed that the coordination 
number ($q$) distributions of the spherical mother boundary surfaces 
of three-dimensional DT mfds are consistent to the theoretical prediction 
for the two-dimensional random surfaces in the large-$q$ region.

Here, we concentrate on the boundary surfaces in three dimensions.
The ensemble of these boundary surfaces consists of various volume 
surfaces. 
The boundary surfaces are divided into two classes: one is a baby 
boundary and the other is a mother one. 
Here, we give a precise definition of ``baby'' and ``mother'' 
universes (see Fig.\ref{Fig:Boundary_3D_CS}). 
The mother universe is defined as a boundary surface 
($\Sigma_{2}$) with the largest volume ($V_{2}$), and 
the other surfaces are defined as baby universes. 
The boundaries with small sizes ($\sim {\cal O}(1)$) are called 
a baby one. 
The baby boundaries originate from the small fluctuations 
of the three-dimensional Euclidean spaces. 
We thus think that these surfaces are non-universal objects. 
In ref.\cite{3D_Scaling} it is reported that 
the surface-area-distributions (SAD) of the mother universe 
show the non-trivial scaling behavior near to the critical point. 
If this scaling behavior turns out to be correct in the strict 
sense of the limit $N \to \infty$, we can take a continuum limit 
of three-dimensional DT mfds as well as two-dimensional DT mfds. 
Therefore, we focus on the mother boundary near to the critical 
point in the following. 
%
\begin{figure}
\centerline{\psfig{file=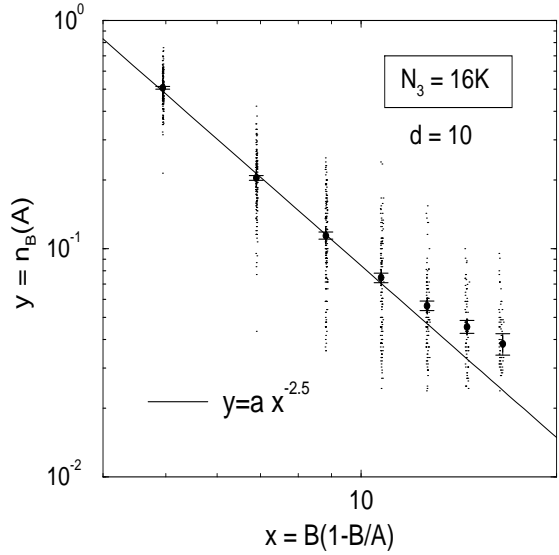,height=8cm,width=8cm}} 
\caption
{
Volume distribution of minbu in terms of the spherical boundary surfaces 
in 3D DT mfds at distance 10.
Total volume of 3D DT mfds is 16K.
The lower limits of the volume of the spherical boundary 
surfaces is 400.
}
\label{Fig:BMinbu_3D_16K_d10}
\end{figure}

%
\begin{figure}
\centerline{\psfig{file=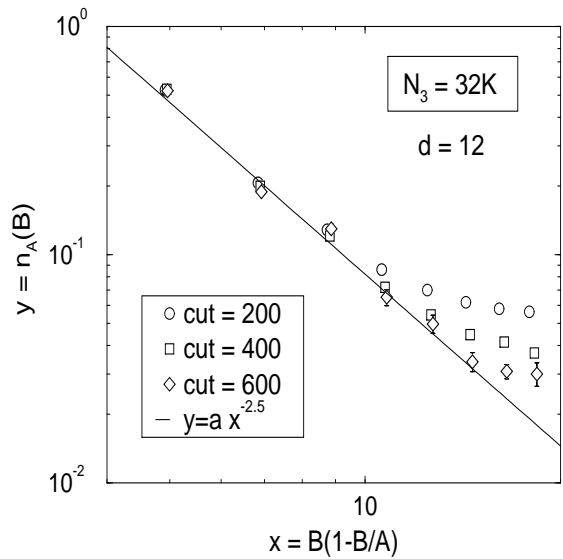,height=8cm,width=8cm}} 
\caption
{
Volume distribution of minbu in terms of the spherical boundary surfaces 
in 3D DT mfds at distance 11.
Total volume of 3D DT mfds is 32K.
There are several lower limits of the volume of the spherical boundary 
surfaces: 200, 400 and 600. 
}
\label{Fig:BMinbu_3D_32K_d12}
\end{figure}

%

The following question now arises: how to measure $\gamma_{st}$ of 
these boundary surfaces with different sizes. 
Suppose that a spherical boundary surface with area $A$ 
is obtained; we can then use the standard minbu algorithm\cite{MINBU1} 
in order to measure $\gamma_{st}$. 
The distribution function for the minbu with size $B$ can 
be written as
\begin{equation}
n_{A}(B) = \frac{3(A-B+1)(B+1)Z[A-B+1]Z[B+1]}{Z[A]}.
\label{eq:Z}
\end{equation}
When we substitute the asymptotic form of the partition function, 
$Z[A] \approx A^{\gamma_{st}-3}e^{\mu A}$, for eq.(\ref{eq:Z}) we 
obtain the normalized distribution function,
\begin{equation}
n_{A}(B) = c \left\{ (B+1)(1-\frac{B-1}{A}) \right\}^{\gamma_{st}-2},
\end{equation}
where c denotes a normalization factor which depends 
on the area ($A$). 
The spherical mother boundaries are selected as grand-canonical 
ensembles. 
In the case that the topology of the boundary surface is not a sphere, 
but a handle-body, such as a torus, the standard minbu algorithm 
does not work well\footnote{
To put it another way, when the topologies of the baby boundary 
and the mother boundary are different, the minbu algorithm 
does not work well.}. 
It is for this reason that we concentrate on the spherical 
boundary surface. 

There are other things to note in terms of the grand-canonical 
ensembles in question. 
Using the grand-canonical ensembles for the measurement 
$\gamma_{st}$ of boundary surfaces, we suffer from 
finite-size effects. 
Therefore, we should introduce a lower limit of the sizes of 
the boundary surfaces. 
Then, the grand-canonical ensemble does not contain surfaces 
whose sizes are less than the lower limit. 
We choose an appropriate geodesic distance corresponding to the total 
volume ($N_{3}$). 

Fig.\ref{Fig:BMinbu_3D_16K_d10} shows the distribution of 
the minbu of the spherical mother boundary surfaces at d$=10$ 
with a total volume ($N_{3}$) of $16K$. 
Fig.\ref{Fig:BMinbu_3D_32K_d12} shows the distribution of 
the minbu of the spherical mother boundary surfaces 
at d$=11$ with $N_{3}=32K$. 
Whether those introduced lower limits are sufficient or not is open 
to question, and the average size of the grand-canonical 
ensembles is about $700$ for lower limit of $600$ 
in Fig.\ref{Fig:BMinbu_3D_32K_d12}. 
We examed separately the string susceptibility exponent 
in two-dimensional pure gravity with a total area of $700$, 
and actually obtained $\gamma_{st} = -0.49(5)$ with the number of 
ensembles being $10^{5}$, even in such a small-size simulation. 
Fig.\ref{Fig:BMinbu_3D_32K_d12} reveals that the scaling property 
of the minbu area distributions becomes more clear the bigger is 
the average size of the boundary. 
Thus, these distributions remain after the thermal limit 
$N_{3} \to \infty$. 
It should be concluded, due to these scaling relations, 
that the string susceptibility exponent of the spherical 
mother boundary surfaces in three dimensions 
is completely consistent with the string susceptibility 
exponent of the random surface in two dimensions. 
Our numerical results given here agree with the discussions in refs.
\cite{ver,ishi}. 

On the other hand, when we apply the method in the previous 
section to higher dimensions, we are confronted by a difficulty.
We cannot use Euler's character in order to distinguish 
the topologies of three-dimensional DT mfds. 
That is to say, to chose a spherical three-dimensional boundary 
mfd in four-dimensional mfd is not easy at all.
Thus, it is difficult to determine $\gamma_{st}$ 
of the boundary mfds in four dimensions by means of the minbu 
technique because of the reason mentioned above. 
It is too involved a subject to be treated here in detail. 

Thus in the following section we consider the scaling 
structures of the minbu distributions in three- and 
four-dimensional simplicial quantum gravity. 
%
\section{Fractal structures of three- and four-dimensional DT mfds}
\begin{figure}
\centerline{\psfig{file=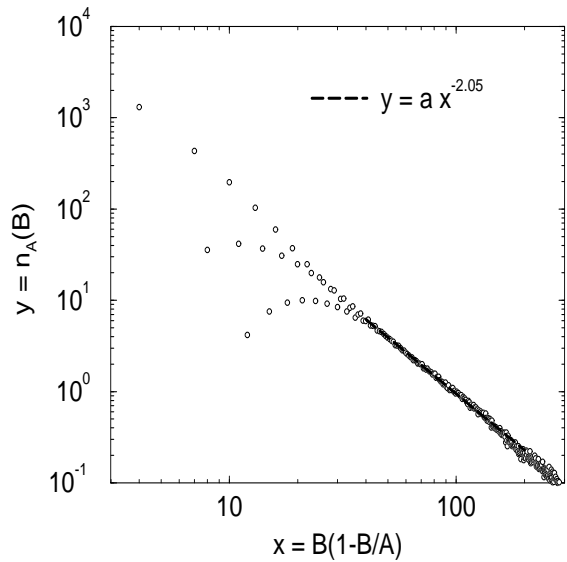,height=8cm,width=8cm}} 
\caption
{
Minbu volume distribution $n_{A}(B)$ of the three-dimensional DT mfds with 
$N_{3}=16K$ with log-log scales.
Coupling constants $\kappa_{0}=4.09$ and $\kappa_{3}=2.20$ for $N_{3}=16K$.
Dotted line is drown using the least squares with the data points of $800$ 
configurations with a range of (x=)$40 \sim 200$.
}
\label{Fig:SMinbu_3D_16K}
\end{figure}

\begin{figure}
\centerline{
\psfig{file=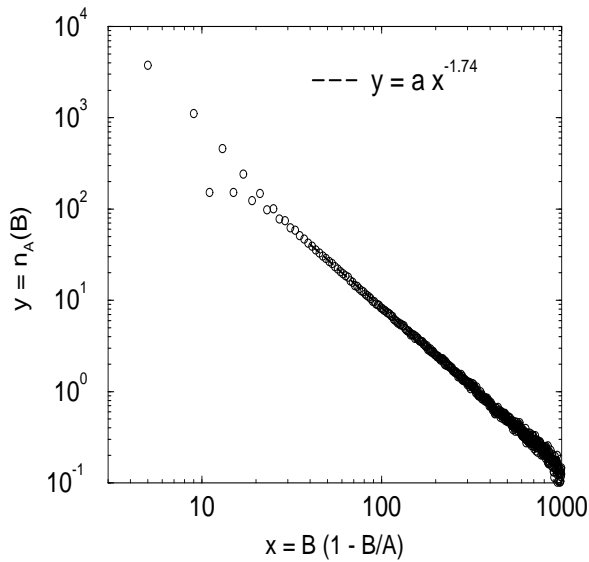,height=8cm,width=8cm}} 
\caption
{
Minbu volume distribution $n_{A}(B)$ of the four-dimensional DT mfds with 
$N_{4}=64K$ with log-log scales.
Coupling constants $\kappa_{2}=1.275$ and $\kappa_{4}=1.353$ for $N_{4}=64K$.
Dotted line is drown using the least squares with the data points of $500$ 
configurations with a range of (x=)$40 \sim 1000$.
}
\label{Fig:SMinbu_4D_64K}
\end{figure}

In the introduction we report on the possibility that this model 
in three- and four-dimensional simplicial quantum gravity 
makes sense near to the critical point belonging to  
the strong-coupling phase.
Therefore, we investigated the scaling structures of 
the minbu distributions near to the critical point belonging to  
the strong coupling phase numerically.
Fig.\ref{Fig:SMinbu_3D_16K} shows the distribution of the minbu 
of three-dimensional DT mfds with a total volume ($N_{3}$) of $16K$.
We tune $\kappa_{0} = 4.09$ and $\kappa_{3} = 2.20$.
The three branches shown in Fig.\ref{Fig:SMinbu_3D_16K} reflect the 
symmetric factor of three-dimensional DT mfds.
The influence of such a symmetric factor has not been observed in 
two-dimensional DT surfaces.
Fig.\ref{Fig:SMinbu_4D_64K} shows the distribution of minbu of 
four-dimensional DT mfds with a total volume ($N_{4}$) of $64K$.
We tune $\kappa_{2} = 1.275$ and $\kappa_{4} = 1.353$.
We also observe two branches which reflect the symmetric factor of 
four-dimensional DT mfds.
In three and four dimensions we have a similar scaling behaviour of the 
minbu volume.
From these scaling date which are shown in Figs.\ref{Fig:SMinbu_3D_16K},
\ref{Fig:SMinbu_4D_64K}, we can extract the string susceptibility 
exponents: $\gamma_{st}=-0.05(6)$ with $N_{3}=16K$ and $\gamma_{st}=0.26(5)$ 
with $N_{4}=64K$\cite{CS_in_Latt97}.
%

%

\subsection{Continuum limit in four-dimensional pure simplicial quantum
gravity}
It is generally agreed that the phase transition of simplicial quantum 
gravity in both three and four dimensions is first 
order\cite{3D_1st,4D_1st}.
Why is the scaling behaviour observed near to the critical point?
Let us devote a little more space to discussing this question.
In ref.\cite{CS_in_4D} a quantum theory of four-dimensional gravity which 
is the analog of the Liouville theory of two-dimensional quantum gravity 
has been argued by considering the effective action for the conformal factor 
of the metric induced by the four-dimensional trace anomaly.
The authors in the reference have also argued for a model including the 
volume (cosmological constant) and Einstein terms.
The string susceptibility exponent ($\gamma_{st}$) obtained by 
ref.\cite{CS_in_4D} in four dimensions has a similar behaviour to the 
$\gamma_{st}$ obtained by the Liouville theory in two dimensions:
\begin{equation}
\gamma_{st}(Q^{2}) = -\frac{Q^{2}}{4} ( 1 - \frac{8}{Q^{2}} + 
\sqrt{1-\frac{8}{Q^{2}}} ) \;\;\; (4 \; \mbox{Dim})
\end{equation}
and 
\begin{equation}
\gamma_{st}(c) = -\frac{1}{12} ( 1-c+\sqrt{(1-c)(25-c)} )
\;\;\; (2 \; \mbox{Dim}).
\end{equation}
Here, $Q^{2}$ which plays the role analogous to matter central charge in 
two dimensions is the coefficient of the Gauss-Bonnet term in the trace 
anomaly\cite{CS_in_4D}. 
If $Q^{2} > 8$ (i.e., $\gamma_{st} (Q^{2}) < 0$, the cosmological and 
the Einstein terms become irrelevant in the infinite volume limit, and 
the spike configurations are suppressed\cite{CS_in_4D}.
Smooth configuration (i.e. smooth phase) is expected to appear.
$Q^{2}=8$ ($\gamma_{st}=0$) behaves in many respects like the $c=1$ case in 
two-dimensional quantum gravity.

\begin{figure}
\centerline{
\psfig{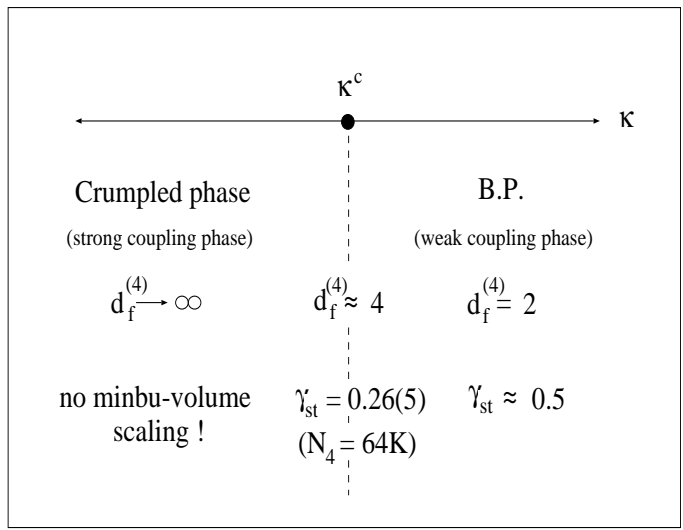}} 
\caption
{
Phase diagram expected from numerical simulations for pure 4D simplicial 
gravity.
$d^{(4)}_{f}$ indicates the fractal dimension obtained in 4D.
}
\label{Fig:Phase_4D}
\end{figure}

Our numerical results mentioned in the previous subsection, as shown in 
Fig.\ref{Fig:Phase_4D}, suggest that there are no signals for the smooth 
phase in which we can take the continuum limit.
%

%
It seems reasonable to suppose that the scaling relations which we obtained 
near to the critical point are pseudo-scaling relations.

\section{Summary and Discussion}
We investigated the statistical properties of boundary surfaces of 
three-dimensional DT mfds with $S^{3}$ topology near to the critical 
point.
We have found positive evidence that the spherical mother boundary surfaces in 
three dimensions are equivalent to two-dimensional spherical random 
surfaces described by the matrix model.
We thus have a conjecture: the mother boundary surfaces in three dimensions 
with any handles are statistically equivalent to two-dimensional random 
surfaces described by the matrix model.
Therefore, if these boundary surfaces in three dimensions can be recognized 
as random surfaces, three-dimensional DT mfds can be reconstructed by the 
direct products of $\Sigma$ (two-dimensional DT surfaces) and $d$ 
(geodesic distance)\footnote{This strategy can straightforward be applied 
to the four-dimensional case if the equivalence between the boundary DT mfds 
in four dimensions and the three-dimensional DT mfds is firmly established.}.
Unfortunately, we cannot use the Euler's character in order to distinguish 
the topologies of three-dimensional DT mfds.
Thus, it is difficult to determine 
$\gamma_{st}$ of the 
boundary mfds in four dimension by means of the minbu technique.
There is room for further investigation.

Furthermore, we investigated the volume distributions of the minbu of 
three- and four-dimensional DT mfd near to the critical point.
Our numerical results of the minbu volume distribution in three and four 
dimensions show a tendency which is also similar to two-dimensional 
case.
If we assume that the partition function in three and four dimensions can be 
written in the same asymptotic form as the two-dimensional case, we obtain 
the string susceptibility exponents: $\gamma_{st}=-0.05(6)$ with $N_{3}=16K$ 
and $\gamma_{st}=0.26(5)$ with $N_{4}=64K$ near to the critical point.

\begin{center}
{\Large Acknowledgements}
\end{center}
We are grateful to H.Kawai, H.Hagura, N.Ishibashi, T.Izubuchi 
and Y.Watabiki for useful discussions and comments.
One of the authors (N.T.) is supported by Research Fellowships 
of the Japan Society for the Promotion of Science 
for Young Scientists.


\end{document}